\documentclass[twocolumn,showpacs,preprintnumbers,nofootinbib,floatfix]{revtex4}
\usepackage{graphicx}% Include figure files
\usepackage{dcolumn}% Align table columns on decimal point
\usepackage{bm}% bold math
\usepackage{amsfonts}
\usepackage{amsthm}
\usepackage{amsmath}
\usepackage{amssymb}
\usepackage{natbib}
\usepackage{epsfig}

\newcommand{\ch}{{\cal H}}

\newcommand{\bea}{\begin{eqnarray}}
\newcommand{\eea}{\end{eqnarray}}
\newcommand{\be}{\begin{equation}}
\newcommand{\ee}{\end{equation}}

\newcommand{\bi}{\begin{itemize}}
\newcommand{\ei}{\end{itemize}}

\newcommand{\cn}{{\cal N}}
\newcommand{\tg}{\tilde{\Gamma}}

\begin{document}

\title{Inflation without Inflaton(s)}
\author{Scott Watson}
\email{watsongs@physics.utoronto.ca} 
\affiliation{Physics Department, University of Toronto,
Toronto M5S 1A7
Ontario, Canada.}
\author{Malcolm J. Perry}
\affiliation{Department of Applied Mathematics and Theoretical Physics,
Centre for Mathematical Sciences,University of Cambridge,
Wilberforce Road,
Cambridge CB3 0WA, England.}
\author{Gordon L. Kane}
\affiliation{Michigan Center for Theoretical Physics
University of Michigan, Ann Arbor, Michigan 48109, USA.}
\author{Fred C. Adams}
\affiliation{Michigan Center for Theoretical Physics
University of Michigan, Ann Arbor, Michigan 48109, USA.}
\date{\today}

\begin{abstract}
We propose a model for early universe cosmology without the need for fundamental scalar fields.  
Cosmic acceleration and phenomenologically viable reheating of the universe results from a series of energy transitions,
where during each transition vacuum energy is converted to thermal radiation.
We show that this `cascading universe' can lead to successful generation of adiabatic density fluctuations and an observable gravity wave spectrum in some cases, where 
in the simplest case it reproduces a spectrum similar to slow-roll models of inflation.  
We also find the model provides a reasonable reheating temperature after inflation ends.  This type of model may also be relevant  for addressing the smallness of the vacuum energy today.
\end{abstract}
\maketitle

%%%%%%%%%%%%%%%%%%%%%%%%%%%%%%%
\section{Introduction}
Despite the simplicity and promising phenomenology of scalar driven, slow-roll inflation, much remains to make the idea theoretically viable.
In particular, vexing issues such as the required flatness of the inflationary potential and the very existence of a fundamental scalar (which must be both extremely light and
weakly interacting) remain elusive  (see however \cite{Adams:1992bn,Freese:1990rb,Arkani-Hamed:2003mz}).  
In recent years, a substantial effort has been invested in understanding how to embed such models in a quantum theory of gravity \cite{Kachru:2003aw,Kachru:2003sx}, and there has also been the suggestion of removing the need for slow-roll completely \cite{Alishahiha:2004eh} (see \cite{Garriga:1999vw} for earlier work).  However, in this paper we will take a different and yet complimentary approach to inflation model building based on fundamental scalars.

As the universe expands and cools the fields and particles affecting the expansion pass through a number of different phases.
In the very early universe many of these transitions may have been inflationary.
In this paper, we propose the idea of a  `cascading universe' by asking: if the universe passed through enough of these transitions, could this provide an adequate alternative to scalar driven inflation?
Our goal will be to examine whether the proposed cascading universe model can satisfy the rigid constraints required of successful inflation model building.
We will postpone the very important question of embedding the model in a fundamental theory (such as string theory) for future work.

In Section \ref{section2}, we present the cascading model and 
obtain the constraints on the decay rate in order for adequate inflation to solve the standard cosmological problems (e.g. horizon and flatness problems).
In Section \ref{natureoftrans}, we address the nature of the transitions and we find that adequate inflation and
successful reheating are possible given certain constraints on the nature of the transitions.
In particular, we find that cascading can proceed via second order or weakly first order phase transitions and at a rate $\Gamma \lesssim H$ 
as might have been anticipated from intuition coming from the graceful exit problem in old inflation.
Working under these assumptions, in Section \ref{section3} we turn to the issue of cosmological perturbations and demonstrate that a nearly scale invariant spectrum of both density and tensor perturbations results in the simplest case of a constant decay rate.  We also find that in the more realistic case of a varying decay rate it may be possible to distinguish this model from the usual slow-roll models in that the evolving adiabatic sound speed can result in an observable tensor to scalar ratio.
We conclude in Section \ref{section4}, where we summarize our results and discuss future considerations. 
\section{A Cascading Universe \label{section2}}

Let us consider the case of a universe dominated by vacuum energy and an additional radiation component that is subdominant.
For a homogeneous and isotropic universe the Einstein equations are
\bea \label{eom1}
3H^2&=&\frac{8 \pi}{M_p^2}  \rho, \nonumber \\
\frac{\ddot{a}}{a}&=& -\frac{4 \pi }{3 M_p^2} \left(  \rho+3 p \right), \nonumber \\
\dot{H}&=&- \frac{4 \pi}{M_p^2}\left( \rho+p \right),
\eea
where we work with the Planck mass, which is related to the Newton constant by $G_N=M_p^{-2}$.
The continuity equation is given by
\bea \label{eom2}
\nabla_\mu T^{\mu \nu}&=&0, \\
\Rightarrow \dot{\rho}&=&-3H\left( \rho+p \right), 
\eea

We consider a two component fluid composed of radiation and vacuum energy.  The energy density and pressure are given by
\bea
\rho=\rho_\Lambda+\rho_r,  \;\;\;\;\;\;\;\;\;\;    p=p_\Lambda+p_r, \\
\rho_\Lambda=\frac{\Lambda M_p^2}{8 \pi},  \;\;\;\;\;\;\;\;\;\;  \rho_r=\frac{\rho_0}{a^4}, \\
p_\Lambda=-\rho_\Lambda,  \;\;\;\;\;\;\;\;\;\; p_r=\frac{1}{3} \rho_r,
\eea
with $\Lambda>0$.
For these sources the equations of motion (\ref{eom1}) and (\ref{eom2}) become
\bea
3H^2&=&\frac{8 \pi}{M_p^2} \left( \rho_\Lambda+\rho_r \right), \nonumber \label{eom3} \\
\frac{\ddot{a}}{a}&=& \frac{8 \pi }{3 M_p^2} \left( \rho_\Lambda - \rho_r \right),\nonumber \\
\dot{H}&=&- \frac{16\pi }{3M_p^2}  \rho_r, \nonumber \\
\dot{\rho_r}&=&-4H\rho_r, 
\eea
where we have used $\dot{\rho_\Lambda}=0$.
We see that in order for acceleration to occur we need $\rho_r<\rho_\Lambda$.  
In fact, the amount of radiation present is a measure of how far the universe is from an exactly de Sitter phase.
Quantitatively this can be seen by considering 
\be
\frac{d}{dt} (H^{-1})=-\frac{\dot{H}}{H^2} \equiv \hat{\epsilon},
\ee
where $\hat{\epsilon}$ is  a parameter measuring the deviation from a pure dS space-time\footnote{This is analogous to the slow-roll parameter
$\epsilon$ in models of scalar field inflation, but because our model does not contain any scalar fields we will avoid this terminology.
Moreover, in contrast to the slow-roll case, the definition of $\hat{\epsilon}$ is exact and does not depend on any approximation.}.
For inflation to occur we thus expect $\hat{\epsilon} \ll 1$, which for this background gives
\be \label{slp}
\hat{\epsilon}=\frac{2 \rho_r}{\rho_\Lambda+\rho_r} \ll 1.
\ee

One can solve the background equations (\ref{eom3}) in the absence of a coupling and we find
\bea
a^2(t)&=&c_0^2 \: \sinh \left(      \sqrt{    \frac{4 \Lambda}{3}    } t +  c_1 \right) , \\
H(t)&=& \sqrt{\frac{\Lambda}{3}}\coth \left(   \sqrt{\frac{4 \Lambda}{3}      } t +  c_1 \right),
\eea
where
\be
c_0^2=\left(\frac{8 \pi \rho_0}{M_p^2  \Lambda} \right)^{1/2} = \frac{1}{\sinh( c_1)},
\ee
are constants chosen so that when $t=0$ we have $a=1$.

Since this solution is derived for the case of $\Lambda$ = {\sl
constant}, inflation of course does not end, and this solution does not provide
for a successful inflationary epoch.  When we include
time varying values of the vacuum energy, which results from multiple transitions or cascades, we will see
the result becomes satisfactory.

The energy transfer from the vacuum energy density to radiation (massless string states)
is given by
\bea \label{qeq}
Q_\Lambda&=&-\Gamma \rho_\Lambda, \nonumber \\
Q_r&=&\Gamma \rho_\Lambda.
\eea
where $\Gamma$ is the transition rate
and the modified continuity equation becomes
\bea
\nabla_\mu T^{\mu 0}_\Lambda &=&\dot{\rho}_\Lambda=Q_\Lambda, \\
\nabla_\mu T^{\mu 0}_r &=&\dot{\rho}_r+4 H \rho_r=Q_r, \\
&\Rightarrow &\nabla_\mu \left( T^{\mu \nu}_\Lambda + T^{\mu \nu}_r \right)=0. \\
\eea
The equations of motion for the background are then
\bea
\label{e1} 3 H^2&=&\frac{8 \pi}{M_p^2} \left( \rho_\Lambda + \rho_r \right), \\
\label{e2} \dot{\rho}_\Lambda&=&-\Gamma \rho_\Lambda, \\
\label{e3} \dot{\rho}_r&=&-4 H \rho_r + \Gamma \rho_\Lambda. 
\eea
From (\ref{e2}) we immediately find 
\be
\rho_\Lambda=\rho_{\Lambda_0} e^{- \Gamma t}.
\ee
Using this result in the above equations we find
\be
\dot{H}+2 H^2 = \frac{16 \pi}{3 M_p^2} \rho_{\Lambda_0} e^{- \Gamma t}.
\ee
The solutions are related to modified Bessel functions.  They can be simply expressed by introducing the dimensionless quantity
\be 
\tau= \sqrt{ \frac{128 \pi \rho_{\Lambda_0}}    {3 M_p^2 \Gamma^2 }  }    \: e^{-{ \Gamma t}/{2}  } \equiv \tau_0 e^{-{ \Gamma t}/{2}  } .
\ee
The scale factor and Hubble parameter are then given by
\bea 
a^2&=&\frac{4}{\Gamma} \left( \alpha_1 I_0(\tau) + \alpha_2 K_0(\tau) \right), \label{thisisasq}\\
H&=& \frac{\Gamma \tau}{4} \left( \frac{\alpha_2 K_1(\tau) -\alpha_1 I_1(\tau) }{\alpha_2 K_0(\tau) + \alpha_1  I_0(\tau) } \right), \label{thisisH}
\eea
where the functions $I_\nu$ and $K_\nu$ are modified Bessel functions of order $\nu$ (see e.g. \cite{stegun}).
The constants are given by 
\bea
\alpha_1&=& \frac{\Gamma \tau_e}{4}  K_1(\tau_e) - H_e K_0(\tau_e)  , \\
\alpha_2&=&  \frac{\Gamma \tau_e}{4} I_1(\tau_e) + H_e I_0(\tau_e),
\eea
with $\tau_e=\tau(t_e)$,  $H_e$ the Hubble parameter at the end of inflation ($t=t_e$) and we normalize so that the number 
of efoldings is measured from the end of inflation,  $N=\ln (a_e/a)=-\ln(a)$ where $a_e=1$. 

%%%%%%%%%%%%%%%%%%%%%%%%%%%%%%%%%%%%%%%
\begin{figure}[t]
\includegraphics[width=8.2cm]{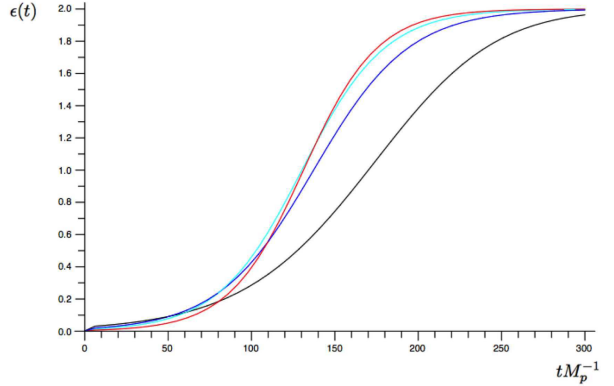}
\caption{\label{fig1} The deformation parameter for various values of the decay rate $\Gamma$.
Time is measured in units of the Planck time and we take $M_p^4 \gg \rho_{\Lambda_0}\gg \rho_r$.
The various curves are given by the values $\Gamma=0.01, 0.02, 0.03, 0.05$ from left to right.
As discussed in the text, $\hat{\epsilon}$ is initially small and proportional to the density of radiation $\rho_r$.  As 
inflation proceeds, more and more energy is dumped into radiation via the coupling $\Gamma$.  
At the very end of inflation we are left with a radiation dominated universe corresponding to $\hat{\epsilon}=2$.}
\end{figure}
\begin{figure}[t]
\includegraphics[width=7.7cm]{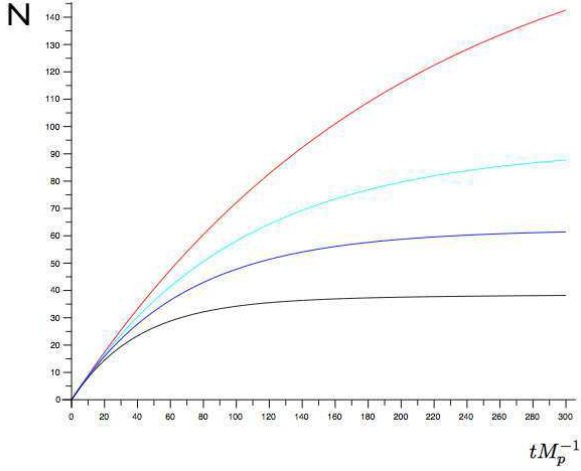}
\caption{\label{fig2} The number of e-folds of inflation $N=\int H dt$ for various values of the decay rate $\Gamma$.
Time is measured in units of the Planck time and we take $M_p^4 \gg \rho_{\Lambda_0}\gg \rho_r$.
The various curves are given by the values $\Gamma=0.01, 0.02, 0.03, 0.05$ from top to bottom.
We see that the requirement of sufficient inflation places a constraint $\Gamma \le 0.02$.}
\end{figure}
\begin{figure}
\includegraphics[width=6.5cm]{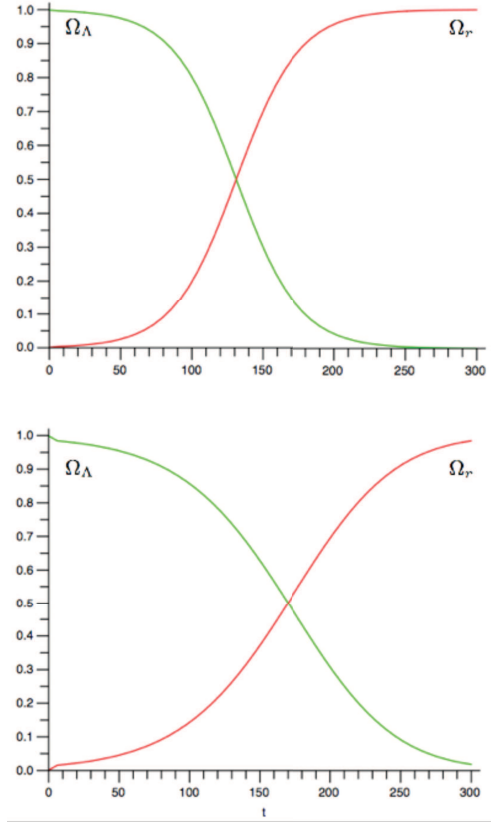}
\caption{\label{fig3} The graphs above show the evolution of the vacuum energy density $\Omega_\Lambda=\rho_\lambda/\rho$ and the radiation energy density 
$\Omega_r=\rho_r/\rho$ relative to the total density.  We present the evolution for two values of the coupling  $\Gamma=0.05$ (top) and $\Gamma=0.01$ (bottom),
where it an be seen that stronger coupling means inflation ends faster,
through faster dissipation.}
\end{figure}

%%%%%%%%%%%%%%%%%%%%%%%%%%%%%%%%%%%%%%%%%%%%%%%%%%%%
We can again introduce a deformation parameter as in (\ref{slp}), however now it is time dependent,
\be \label{eps1}
\hat{\epsilon}(t)=\frac{2 \rho_r}{\rho_\Lambda+\rho_r}=2 \Omega_r.
\ee
Using this in (\ref{e1}) we find
\be
\hat{\epsilon}(t)=2 - \frac{16 \pi \rho_{\Lambda_0}}{3 H^2 M_p^2}e^{-\Gamma t},
\ee
with $H$ being given by (\ref{thisisH}).
At the beginning of inflation we have $H^2 M_p^2 \sim \rho_\Lambda$ and so $\hat{\epsilon} \ll 1$.
As the energy is transferred from the vacuum energy density to radiation via particle production, the deformation parameter increases as can be seen in Fig. \ref{fig1}.
Inflation then ends when $\rho_r \approx \rho_\Lambda$ and $\hat{\epsilon} \approx 1$, which can be seen in Figures \ref{fig1} and \ref{fig3}.
The final result is $\hat{\epsilon} =2$ and we are left with a universe filled by the radiation\footnote{Much later, of course, the
radiation is diluted as the volume increases and the small remaining
constant energy density again dominates.} $\rho_r$.
Let us now consider the amount of inflation or number of e-folds.
Using the above expression for the deformation parameter the Hubble equation can be rewritten as
\be
3 H^2=\frac{16 \pi \rho_{\Lambda_0} }{M_p^2(2-\hat{\epsilon})} e^{-\Gamma t /2}.
\ee  
%%%%%%%%%%%%%%%%%%%%%%%%%%%%%%%%%%%%%%%%%%%%%%%%%%%%
%%%%%%%%%%%%%%%%%%%%%%%%%%%%%%%%%%%%%%%
This can then be integrated to find the number of efoldings,
\bea  \label{efoldconst}
N&=& \left(     \frac{16 \pi \rho_{\Lambda_0}}{3 M_p^2}      \right)^{1/2}      \int^{t_0}_{t_e}       \frac{      e^{-\Gamma t/2}    }{    \sqrt{2-\hat{\epsilon}}   } \, dt, \nonumber \\
&\approx& \frac{2}{\Gamma}\left(  \frac{16 \pi \rho_{\Lambda_0}}{3 M_p^2}\right)^{1/2}  \sim \frac{8 \rho_{\Lambda_0}^{1/2}}{\Gamma M_p} 
\sim 8 \frac{ H_b}{\Gamma},
\label{efoldsf}
\eea
where $t_0=0$ is the beginning of inflation, $t_e$ is the end and $H_b$ is the initial Hubble scale.
In the second line we use the fact that the denominator varies smoothly from $\sqrt{2}$ to $1$ and 
$\exp(-\Gamma t_e/2) \approx 0$.  As we may have anticipated the amount of inflation depends on the initial vacuum density and the decay rate.

As an example, consider inflation with a Hubble scale near the GUT scale $H_b \sim M_{GUT} \approx 10^{15} \, GeV$.  We see to get adequate inflation the decay rate need only be 
slightly below the initial Hubble scale $\Gamma \sim  10^{14} \, GeV$.   This condition is required in order that cascading lasts long enough so that the cosmic acceleration can resolve the horizon and flatness problems.
In Figures 1-3, we examine the evolution numerically and find adequate inflation is possible given modest
values of the parameters.

Another important consideration is the reheat temperature of the model.
The cosmic acceleration ends at the moment $t_r$ when $\rho_r=\rho_\Lambda$ and radiation comes to dominate.
At this moment we have $3 H^2 = 16 \pi \rho_\Lambda$ where 
$\rho_\Lambda =\rho_r= \rho_{\Lambda_0}  e^{-\Gamma t_r}$.
Using the exact solution (\ref{thisisasq}) and (\ref{thisisH}) and assuming the minimal amount of efoldings ($N=60$) we find 
$\Gamma t_r \approx 10$ so that 
the reheating temperature can 
be approximated as
\bea
T_r &\approx& \rho_r^{1/4} =  \rho_{\Lambda_0}^{1/4} e^{-\Gamma t_r/4}, \nonumber \\
&\approx&     10^{15} \, GeV,
\eea
where we have used $\rho_{\Lambda_0} = {\Lambda_0 M_p^2}/{8 \pi}$ and we have taken the initial Hubble scale $H_b \approx 10^{14} \, GeV$.
We will see in the next section that this is consistent with producing the 
observed temperature anisotropies in the cosmic microwave background and avoiding over production of gravity waves.

\section{Nature of the Transitions \label{natureoftrans}}
Another important constraint on the cascading model comes from considering the type of phase transition from level to level.
In arriving at the previous constraint on the decay rate $\Gamma$ in the last section, we have tacitly assumed that whatever the nature of the phase transition that it was successfully completed and that the associated microphysics was irrelevant. However, obviously this is not always the case and properties of the transitions, such as whether they are first or second order can play an important role.  We will now consider both cases of first and second order transitions.

First order transitions typically proceed by nucleation of bubbles of the new phase in the background of the old phase (see however \cite{Witten:1984rs}).
The energy difference between the phases is stored in the bubble walls, and typical expand near the speed of light.
The transition from the old phase to the new phase is complete when all the nucleated bubbles collide, releasing the energy stored in their walls.

Completion of such transitions in cosmological backgrounds can often be problematic.
If the tunneling barrier (more precisely, the tunneling action) is large, then this corresponds to a strongly first order transition.
In such cases the average bubble size is typically comparable to the gravitational scale ($H^{-1}$) and the gravitational background can have important 
effects \cite{Coleman:1980aw,Guth:1982pn,Kamionkowski:1993fg}.
In particular, in the case of inflationary backgrounds, bubbles of the new phase will form in the exponentially expanding background of the old phase.
In this case, although the nucleated bubbles expand at the speed of light, the background itself is expanding faster.  This makes bubble collisions rare and instead of 
completion of the phase transition we find isolated bubbles expanding in a background of eternal inflation.  This is the graceful exit problem.

This was made more precise by Guth and Weinberg in \cite{Guth:1982pn}.  Consider the zero temperature bubble nucleation rate per volume\footnote{Note this is not the same as the decay rate $\Gamma$ that we have introduced above, however we will see that these quantities can be related. } 
\be \label{gam}
\tg = A e^{-S_E},
\ee
where $S_E$ is the Euclidean action and  $A$ comes from a one-loop determinant factor that depends on the microphysics and is typically the energy (density) scale of the transition. We note with hindsight that if the transition is going to complete then the typical bubble size $r_b$ must be much less than the gravitational scale\footnote{A detailed discussion of this point appears in \cite{Kamionkowski:1993fg}, where it is shown that the only feasible first order transition is one that results in a distribution of bubble sizes sharply peaked around $r_b$ and with $r_b \ll H^{-1}$, i.e. a weakly first order transition.}, i.e. $r_b \ll H^{-1}$. In this case gravitational effects in (\ref{gam}) are negligible.  The authors of \cite{Guth:1982pn} then showed that the probability of a point to remain in the false vacuum is given by
\be
p(t) \sim \exp \left(  - \frac{4 \pi}{3} \beta H t \right),
\ee
where we introduce the dimensionless quantity $\beta =\tg / H^4$.  We see that the corresponding decay time is given by
\be
\tau= \left( \frac{4 \pi \beta H}{3}\right)^{-1} = \frac{3}{4 \pi} \frac{H^3}{\tg}.
\ee
The number of efoldings resulting from the transition is $N \approx H \tau$.  The authors of \cite{Guth:1982pn,Kamionkowski:1993fg} found that for percolation to occur and the transition to complete $\beta \gtrsim \beta_c=9/(4 \pi )$,  which we see corresponds to $N \lesssim 1/3$. Thus, we see in the cascading model we have presented in this paper, that if we assume transitions are first order than they must be weakly so and the maximum number of efoldings per level is $N=1/3$ corresponding to $\tg/H^4 = 9 / (4 \pi )$.  To build a successful inflation model we will need a large number of transitions  ( $\approx 150-180$) of this type to get the required $50-60$ efoldings in order to solve the horizon and flatness problems.
This idea for the case of a fundamental scalar field has been argued for in \cite{Freese:2004vs,Freese:2005kt} and is similar to the original proposal of Abbott \cite{Abbott:1984qf}.
To relate this to the phenomenological cascading model we presented in the last section, we note that this is exactly what one expects from a course grained approach.  The decay of vacuum energy to radiation will proceed as explained above, but on gravitational scales $t \gtrsim H^{-1}$ this will simply be seen as a decay of the vacuum energy to radiation, whereas on very small scales $t \ll H^{-1}$ one would see that it was small bubble collisions that had been responsible for replenishing the radiation bath. Thus, by treating the percolation events as instantaneous (as seen from the gravitation scale) the constraint on $\Gamma$ (as opposed to $\tg$) is simply that of the last section $\Gamma/H \lesssim 1/8$.  
Combining both the microscopic constraints, along with the macroscopic requirements we see that the first order transition can work, but it is in a very special regime.
It will be a challenge to embed such an approach into a fundamental theory, such as string theory, where e.g. in the case of scalars, potential barriers typically lead to strongly first order transitions.

Of course the other possibility is that the transitions proceed via a second order phase transition.
Since such transitions do not proceed via quantum mechanical tunneling, bubble percolation is no longer a concern.
Immediate examples of such transitions are provided by that of a scalar field, in which case we simply recover slow-roll inflation models such as new inflation and chaotic inflation.  
Another possible example is provided by that of the `thermalons' \cite{Gomberoff:2003zh}, where transitions are stimulated by thermal fluctuations and so-called `over-barrier' tunneling.

In summary, we find that the cosmological phase transitions of the cascading model must either be second order or weakly first order.

\section{Cosmological Perturbations \label{section3}}
%%%%%%%%%%%%%%%%%%%%%%%%%%%%%%%%%
Let us consider density and tensor fluctuations about the background solution (\ref{eom3}).
We will be primarily interested in modes that leave the Hubble radius $50 - 60$ e-folds before the end of inflation, since these are 
the modes that are responsible for the CMBR anisotropies observed today.
Therefore, instead of working with the exact solution (\ref{eom3}) for the study of perturbations, it will often be simpler to work in the conformal time
$\eta= -\infty \ldots 0$ where $d \eta = a^{-1} dt$ and with approximate solution
\bea
a(\eta)&=& (-\eta)^{-(1+\hat{\epsilon})}, \\
\ch(\eta)&=& \frac{1+\hat{\epsilon}}{-\eta},
\eea
where $\ch$ is the conformal Hubble parameter $\ch=aH$ and is related to the deformation parameter by
$\hat{\epsilon}=1-\frac{\ch^\prime}{\ch^2}$.  The approximate solution treats the deformation parameter as a constant, since its
rate of change is small ($ \dot{\hat{\epsilon}} \sim \hat{\epsilon}$), but of course this solution must break down towards the end of inflation when $\hat{\epsilon} \sim 1$.

We now consider linearized perturbations about the background (\ref{eom3}) and in 
what follows we will adopt the conventions of \cite{Mukhanov:1990me}.

We work in longitudinal gauge with the perturbed line element
\be
ds^2=-(1+2\Phi) dt^2 + (1+2 \Psi + h_{ij}) dx^i dx^j,
\ee
where the tensor perturbation is traceless and transverse (i.e., $h_i^i=\nabla_ih^{ij}=0$)  and
can be broken into its two polarizations $h_\pm$.
Our background contains no anisotropic stress so one finds from the Einstein equations $\Phi=\Psi$.
Thus, we have only one scalar metric degree of freedom associated with the density perturbation and two tensor metric degrees of freedom for the gravity waves.
Because we are considering linearized perturbations the scalar and tensor metric fluctuations decouple and we will treat 
each one in-turn.

\subsection{Density Fluctuations}
Working in conformal time the equations for density perturbations are,
\bea
\nabla^2 \Phi-3 \ch(\ch \Phi + \Phi^\prime)  &=& 4 \pi G a^2 \delta \rho \label{peom1} \\ %constriant
\partial_i \left(  \ch \Phi + \Phi^\prime \right) &=& 4 \pi G a (\rho + p) \delta u_i, \;\;\;\;\;\;\;\;\; \label{peom2} \\    %constraint
\Phi^{\prime \prime}  + 3 \ch \Phi^\prime +(2 \ch^\prime +\ch^2) \Phi &=& 4 \pi G a^2 \delta p, \;   \label{peom3} % dynamic equation
\eea
where $\delta \rho$, $\delta p$, and $\delta u_i$ are the perturbations of the total energy density, pressure, and velocity, respectively and 
$\ch= \frac{d a}{d \eta} a^{-1}$ is the conformal Hubble parameter and $\nabla$ is the comoving gradient.
The pressure perturbation is related to the energy density and entropy density perturbations by
\bea \label{perteomc}
\delta p&=&\left. \frac{\partial p}{\partial \rho} \right\vert_{s}\delta \rho+ \left. \frac{\partial p }{\partial s}\right\vert_{\rho} \delta S, \\
&=&c_s^2 \delta \rho+\tau \delta S, \label{tdst}
\eea
with $c_s$ the adiabatic sound speed at which the perturbations evolve and
should not be confused with the equation of state parameter $w=p/\rho$, which depends on the background quantities.

Indeed, for the model we consider here these quantities are quite different.
The adiabatic sound speed for long-wavelength perturbations is given by\footnote{Strictly speaking it is incorrect to think of $c_s^2 = \frac{\dot{p}}{\dot{\rho}}$ 
as the sound speed during inflation, since this is only true on large scales where $c_s^2=\left. \frac{\delta p}{\delta \rho} \right\vert_{S}\approx \frac{\dot{p}}{\dot{\rho}}$
and these modes evolve on scales beyond the sound horizon.  On small scales during inflation the metric perturbation $\Phi$ oscillates and the effective adiabatic sound speed is found to be $c_s^2=1$, in agreement with causality.}
\be \label{ss}
c_s^2=\frac{\dot{p}}{\dot{\rho}}=\frac{1}{3}-\frac{\Gamma \rho_\Lambda}{3 H \rho_r},
\ee
where we see in the limit $\Gamma \rightarrow 0$ our perturbations will evolve like pure radiation.
This is consistent with the fact that a true cosmological constant does not propagate, i.e. is constant.
This allows us to see the importance of the graviton and other particle production ($\Gamma$ term), 
since in an quasi-exponentially expanding background 
with no transfer the perturbations will be immediately damped away.

The equation of state parameter is given by
\bea
w&=&\frac{p}{\rho}=-1+\frac{4 \rho_r}{3(\rho_\Lambda+\rho_r)}, \\
&=&-1+\frac{2}{3} \hat{\epsilon},
\eea
which reduces to the pure de Sitter solution if $\rho_r =\hat{\epsilon}=0$ as we have noted.  We see that $w(t)$ is explicitly time dependent
and in general this implies that there can be a significant contribution from non-adiabatic pressure in (\ref{tdst}) by the entropy term $\tau \delta S$.  This could result in significant generation of entropy perturbations, a possibility that we will analyze in Section (\ref{thissec}).

We now return to solving the system (\ref{peom1})-(\ref{peom3}).
Combining equation (\ref{peom1}) with (\ref{peom3}) and working in momentum space $\nabla^2 \Phi \rightarrow -k^2 \Phi$, we find a second order 
differential equation,
\bea \label{perteqnc}
\Phi_k^{\prime \prime} &+& 3 \ch (1+c_s^2) \Phi_k^\prime +\left[ c_s^2 k^2 +2 \ch^\prime +\right. \\ &+& \left. (1+3c_s^2)\ch^2  \right] \Phi_k=4 \pi G a^2 \tau \delta S.
\eea
subject to the constraint (\ref{peom2}).
We can simplify this equation by introducing the field redefinition,
\bea
u_k&=& \frac{\Phi_k}{4 \pi G \rho^{1/2} \sqrt{1+w}}, \\
\theta^2 &=& \frac{8 \pi M_p^{-2}}{3 a^2 (1+w)},
\eea
where $w=p / \rho$.
Then (\ref{perteqnc}) becomes
\be \label{ueom}
u^{\prime \prime}_k+ \left( k^2 c_s^2 - \frac{\theta^{\prime \prime}}{\theta} \right) u_k= {\cal N}
\ee
with ${\cal N}$ giving the contribution from entropy modes as
\be \label{theN}
{\cal N}=a^2 \rho^{1/2}\sqrt{ 1+w} \;  \tau \delta S.
\ee

\subsubsection{Adiabatic Fluctuations}
We will first consider solutions to (\ref{ueom}) in the absence of entropy modes, i.e. $\cn=0$.
For modes that are far inside the sound horizon $kc_s \gg \ch$ we find that $u$ oscillates with a constant amplitude
that is to be found by the initial conditions after quantization.
Far outside the horizon $k c_s \ll \ch$ and by inspection we have the solution $u \sim \theta$.  
However, this solution corresponds to a decaying mode for the metric perturbation $\Phi$.
Instead, it is the growing mode that is of interest, which can be found by noting $\theta^{\prime \prime} \approx 0$ during inflation, so that $u$ constant is also a solution.
Assuming that we are deep in the inflation epoch where $\hat{\epsilon}$ is nearly constant an exact solution can be found by integration.  In terms of the original 
metric perturbation $\Phi$ one finds
\bea
\Phi&=&\frac{\ch}{a^2} \left( A_1 + A_2 \int \left( 1+w \right) a^2 d\eta \right), \\
&=&\frac{\ch}{a^2} \left(  A_1 + \frac{2}{3} A_2 \int  \hat{\epsilon} a^2 d\eta \right), \\
&=& A_1 \frac{\ch}{a^2}  + \Phi_0 \hat{\epsilon},  \label{terms1}
\eea
where we have used $w=-1 + \frac{2}{3} \hat{\epsilon}$ is nearly constant during inflation.
The first term in (\ref{terms1}) corresponds to the decaying mode found above ($u \sim \theta$), whereas the second mode is nearly constant with $\Phi_0=\frac{2}{3}A_2$
to be determined by matching to the oscillating mode inside the sound horizon.  
We note the importance of the graviton production in this model resulting in a non-zero radiation density, since 
in the pure de-Sitter case where $\rho_r=0$ so that $\hat{\epsilon}=0$ we see no density metric perturbation would remain since $\Phi \rightarrow 0$ as $\hat{\epsilon} \rightarrow 0$
and all that is left is the decaying mode.  This is an illustration of the no hair theorem for pure de Sitter space.

In summary, we have found that during inflation the metric perturbation is nearly constant on super-horizon scales, whereas on sub-horizon scales
we find $u \sim \Phi$ undergoes constant amplitude oscillations.  What remains is to quantize the perturbations in order to determine the unknown constant $\Phi_0$.
However, we must first justify neglecting the entropy mode term ( i.e. ${\cal N}$) in (\ref{ueom}). 

\subsubsection{Entropy Flucutations \label{thissec}}
In this section we consider the role of entropy fluctuations in the model.
We will follow \cite{Malik:2002jb} where a systematic procedure for the study of perturbations in multi-fluid systems was described.
It will be useful to introduce $\zeta$, which is curvature perturbation on
constant energy density hypersurfaces. We will drop the momentum index in what follows, writing $\zeta \equiv \zeta_k$.  

In the presence of multiple fluids, the total curvature perturbation can be expressed
as a sum of the curvature perturbation due to each fluid component as
\be
\zeta=\sum_\alpha \frac{{\rho}^\prime_\alpha}{{\rho}^\prime} \zeta_\alpha,
\ee
where
\be
\zeta_\alpha=\Phi + \frac{\ch }{{\rho}^\prime_\alpha}\delta \rho_\alpha,
\ee
and we have used the lack of anisotropic stress to again write $\Phi=\Psi$ as before.
For entropy fluctuations we are interested in the non-adiabatic contribution to the pressure perturbation in (\ref{tdst}), which is given by
\bea
\delta p_{nad} &\equiv& \tau \delta S= \delta p - c_s^2 \delta \rho.
\eea
As discussed in \cite{Malik:2002jb}, there are two sources of non-adiabatic pressure
\be
\delta p_{nad}=\delta p_{nad}^{rel} + \delta p_{nad}^{int},
\ee
which are the relative and intrinsic non-adiabatic pressures, respectively.  In the model we are considering here the two fluid components have fixed equation of state, i.e. $\delta p_\Lambda = - \delta \rho_\Lambda$ and $ \delta p_r = 1/3 \delta \rho_r$ so that there is no intrinsic non-adiabatic pressure, i.e. $\delta p^{int}_{nad}=0$.  The contribution to the relative non-adiabatic pressure is \cite{Malik:2002jb}
\bea \label{nadp}
\delta p_{nad}^{rel}&=& -\frac{1}{6 { H} \dot{\rho}} \sum_{\alpha, \beta} \dot{\rho}_\alpha \dot{\rho}_\beta (c_\alpha^2 - c_\beta^2) S_{\alpha \beta},
\nonumber \\  &=& -\frac{1}{3 {\cal H}\dot{\rho}} \dot{\rho}_\Lambda \dot{\rho}_r (c_\Lambda^2 - c_r^2) S_{\Lambda r},
\eea
where we have introduced the relative entropy perturbation
\bea \label{rep}
S_{\alpha \beta}&=&3(\zeta_\alpha-\zeta_\beta), \;\;\;\;\;\ \\
&=&-3H \left(  \frac{\delta {\rho}_\alpha}{\dot{\rho}_{\alpha}} -\frac{\delta \rho_\beta}{{\rho}^\prime_\beta} \right),
\eea
with the factor of three due to the convention of normalizing to baryons.
For the model we consider here
\be
S_{\Lambda r}= - S_{r \Lambda} =-3H \left( \frac{\delta \rho_\Lambda}{\dot{\rho}_{\Lambda}} -\frac{\delta \rho_r}{\dot{\rho}_r}  \right).
\ee
Returning to (\ref{nadp}) we see that the relative non-adiabatic pressure is proportional to $\dot{\rho}_r$, the rate of change of the radiation density.
During inflation one finds from the background solution (\ref{thisisasq}) and (\ref{thisisH}) that $\dot{\rho_r} \approx 0$.  That is, the transfer of vacuum energy to the radiation density via the coupling $\Gamma$ is just enough to counter the dilution of the radiation by the exponential expansion.
Thus, the non-adiabatic pressure is negligible and we need not worry about the presence of entropy perturbations during inflation.

However, there is a more fundamental reason to expect entropy perturbations to be absent from this model.
The crucial point is that a relative entropy perturbation is produced when two fluids generate different curvature perturbations.
This difference can then be mediated from one fluid to the other via the gravitational background.  
A well known example is the perturbation in the baryon-photon ratio
\be
S_{B \gamma}=3(\zeta_B - \zeta_\gamma)=\frac{\delta \rho_B}{\rho_B} - \frac{3}{4} \frac{\delta \rho_\gamma}{\rho_\gamma},
\ee
which does not vanish because the two fluids are perturbed differently.

However, in the case we consider here things are different.  In the absence of the coupling $\Gamma$ there is only one fluid with propagating 
fluctuations, namely the radiation density with fluctuations $\delta \rho_r$.  In this case the long-wave fluctuations propagate at $c_s^2=1/3$ and the cosmological constant remains a
constant, i.e. $\delta \rho_\Lambda=0$.  In the presence of the coupling $\Gamma$ the fluctuations now propagate at a different adiabatic sound speed 
(\ref{ss}), 
but the two fluids are coupled through their equations (\ref{e2}) and (\ref{e3}) through the term $\pm \Gamma \rho_\Lambda$.  Thus, 
there is really only one propagating degree of freedom and the two fluids do not evolve independently, resulting in $S_{\Lambda r}=0$.

In fact, in this regard this is not unlike the case of inflation by a single scalar field where it is known that there are only adiabatic perturbations.
Instead of working with the scalar directly, we could consider two fluids, one representing the kinetic energy with a stiff equation of state $p_1=\rho_1=1/2 \dot{\phi}^2$ and a second fluid composed of the potential $p_2=-\rho_2=V(\phi)$.  Insisting on this two-fluid description and demanding that the full equations of motion are satisfied we are led to an energy exchange term $Q_{\pm}=\pm \dot{\phi} V^\prime(\phi)$, similar to the case we have above.
However, since we know there is only one degree of freedom, we certainly know that there are no entropy perturbations and no non-adiabatic pressures.  
This can be seen by examining the perturbation equations in full detail, and in particular one finds that the two fluids do not evolve independently due to the coupling $Q_{\pm}$ and the fact that the second fluid does not propagate in the absence of the coupling (i.e. $\delta \rho_2 =0$ for $Q=0$).

In sum, we see that entropy perturbations in the cascading model are negligible during inflation for the case of a constant decay rate $\Gamma$.
For the case of a time varying $\Gamma$, this issue must be revisited, which is work in progress.

\subsection{Spectrum of Fluctuations  \label{iso}}
Having shown that entropy perturbations are negligible, we proceed to find the spectrum of the density fluctuations.
In order to find the power spectrum all that remains is to determine the unknown constant $\Phi_0$ in (\ref{terms1}).
We can then find the gauge invariant, comoving curvature perturbation 
\be \label{curvpert}
{\cal R}_k=\Phi_k + \frac{2}{3} \left( \frac{\Phi_k^\prime + \ch \Phi_k}{1+w} \right),
\ee 
which is related to the curvature perturbation $\zeta_k$ from the last section by
\be
{\cal R}_k=\zeta_k+\frac{1}{3}\frac{k^2 \Phi_k}{\ch^\prime-\ch^2},
\ee
so that for large scales modes (which are the ones of interest) $k \rightarrow 0$ and ${\cal R}_k \rightarrow \zeta_k$.
The density power spectrum is then defined as
\be \label{ps}
{\cal P}_{\zeta}=\frac{k^3}{2 \pi^2} |{\zeta}_k|^2,
\ee
which can be compared with observations.

Finding the constant $\Phi_0$ is accomplished by enforcing the correct initial condition on the modes.
However, these modes are born in their vacuum state far below the Hubble radius.
This requires us to quantize the perturbations, starting their evolution in the standard adiabatic vacuum.
Then we have seen that the solution inside ($k c_s \gg \ch$) oscillates with constant amplitude until Hubble radius crossing where it can be 
matched to the solution outside ($k c_s \ll \ch$)
providing us with the required normalization constant.

The only obstacle to quantization is finding the canonical field which diagonalizes the action.  For the case of hydrodynamical fluids, as we consider here, 
this was done in \cite{Mukhanov:1990me}.
There it was found that the canonical field $v_k$ (the so-called Mukhanov variable) which is related to $u_k$ by
\be
u_k=-\frac{   \left( v_k \theta \right)^\prime        }{   c_s k^2 \theta   } .
\ee
and the curvature perturbation (\ref{curvpert}) by $v_k=z {\zeta}_k$,
reduces the action to that of a harmonic oscillator with time dependent frequency.
In terms of this variable the equation of motion (\ref{ueom}) becomes
\be \label{veom}
v_k^{\prime \prime} + \left( k^2 c_s^2 - \frac{z^{\prime \prime }}{z}    \right) v_k =0,
\ee
where $z=(c_s  \theta)^{-1}$.
In terms of the new variable $v_k$ the solutions on large scales ($k c_s \gg \ch$) are given by $v_k \sim z$.
Notice this is the growing mode of interest in contrast to the classical case where $u_k \sim \theta$  decayed
and it is this squeezing of the quantum state that will result in classical fluctuations on large scales.
On small scales the momentum term dominates and we again have oscillations with constant amplitude.

We could now proceed with the approximate solution, however in the case $\hat{\epsilon} \ll 1$ we can solve (\ref{veom}) exactly.
We find $\frac{z^{\prime \prime}}{z}=\frac{\nu^2-1/4}{\eta^2}$ where $\nu=3/2+\hat{\epsilon}$ and the solutions can be expressed in terms of Hankel functions.
We require that the modes begin in the adiabatic vacuum, which amounts to the condition 
\be
v_k = \frac{1}{\sqrt{2 c_s k}} e^{-i k c_s \eta} \;\;\;\;\; \text{as} \;\; k c_s \eta \rightarrow -  \infty.
\ee
The appropriate solution is then given by
\be
v_k(\eta)= \frac{\alpha}{2} \sqrt{-\eta} \, H^{(1)}_\nu(-k c_s \eta),
\ee
where $H_\nu^{(1)}$ is a Hankel function of the first kind and $|\alpha|=1$.
We can then immediately find the curvature perturbation
\be
{\zeta}_k = \left|  \frac{v_k}{z}  \right|=\frac{1}{2 z} \sqrt{-\eta} \, H^{(1)}_\nu(-k c_s \eta)
\ee
On large scales using the asymptotic expansion of the Hankel function we have
\be
\left| {\zeta}_k \right| \approx \frac{1}{z\sqrt{ 2 \pi k c_s }} \left(  -k c_s \eta \right)^{1/2-\nu},
\ee
and using $z=(c_s \theta)^{-1}$ the power spectrum (\ref{ps}) is
\be \label{scalarpower}
P_{\zeta}=\frac{1}{4 \pi^2 c_s \hat{\epsilon}} \left(  \frac{H}{M_p} \right)^2 \left( -k c_s \eta \right)^{-2 \hat{\epsilon}}.
\ee
We note that this reduces to the standard slow-roll inflation result for the case $|c_s|=1$.
 
The tilt of the power spectrum is given by
\bea
n_s&=&1+\frac{d \ln {\cal P}_{\zeta}}{d \ln k}=1-2 \hat{\epsilon}, \\
&=& 1-4 \Omega_r,
\eea
where $\Omega_r=\rho_r/\rho$.
By noting that $\dot{\rho_r} \approx 0$ during the time modes of interest exit the Hubble radius (i.e. $N \sim 50$), we see that the tilt of 
the spectrum is set by the initial abundance of radiation since $\hat{\epsilon} \approx 2 \Omega_{r_0}$ is constant during inflation.

Comparing (\ref{scalarpower}) to the best fit WMAP3 data \cite{Spergel:2006hy},
\be
P_\zeta = 19.9^{+1.3}_{-1.8} \times 10^{-10}, 
\ee
we find that 
\be
\frac{H}{ c_s \hat{\epsilon}^{1/2}}   \lesssim 10^{-5} M_p.
\ee
Since $\hat{\epsilon} \ll 1$ during inflation this implies an upper bound on the Hubble scale during inflation 
$H \lesssim 10^{14} \; {\text GeV}$.  Combining this with the constraint for adequate inflation from (\ref{efoldsf}), i.e. $\Gamma/H \lesssim 1/N$
we find an upper bound
on the decay rate of the vacuum energy $\Gamma \lesssim 10^{13} \; {\text GeV}$
consistent with our earlier results and our general approach.

\subsubsection{Gravity Waves}
The gravitational wave spectrum is found in much the same way as the spectrum of density perturbations.
One first decomposes the graviton into its two polarizations $h_k^{(+)}$ and $h_k^{(-)}$.  The modes then obey the same equation (\ref{veom}) as the density fluctuations 
except in this case we have $v_k=ah_kM_p$ (where we set $ h\equiv h_{\pm}$) and  $z$ is replaced by the scale factor $a$.
The solution is again in terms of Hankel functions, and one finds for the long-wavelegnth fluctuations
\be
h_k \approx \frac{1}{a \sqrt{ 2 \pi k }} \left(  -k \eta \right)^{1/2-\nu},
\ee
and the power spectrum is
\be
P_h=\frac{8}{\pi^2 } \left(  \frac{H}{M_p} \right)^2 \left( -k \eta \right)^{-2 \hat{\epsilon}},
\ee
with the tilt of the tensor spectrum $n_T=-2 \hat{\epsilon}$.
Thus, we see that the main difference between the tensor and density spectrum is the deformation factor and the presence of the adiabatic sound speed 
in the spectrum of density perturbations.  Our tensor to `scalar' ratio is then
\be
r=\frac{P_h}{P_{\zeta}}= 16 \hat{\epsilon} c_s,   \;\;\;\;\;  \text{Scalar Free Model}
\ee
which contains the adiabatic sound speed $c_s$ evaluated at the time of Hubble radius crossing.  

This is an important result and is similar to 
models of kinetic inflation \cite{Garriga:1999vw}, where the adiabatic sound speed offers a way to distinguish this model
from standard slow-roll inflation which gives instead
\be
r=  16 {\epsilon},   \;\;\;\;\;  {\text{Scalar Slow-roll Inflation}},
\ee
where we recall that $\epsilon \approx \hat{\epsilon}$ is the usual slow-roll parameter which measures the slope of the scalar field potential in units of the Hubble scale.

We have seen the adiabatic sound speed does not differ greatly from the usual slow-roll inflation case for the choice of $\Gamma=constant$
that we have considered here.  However, for the case of non-constant $\Gamma$ this could dramatically change, since the adiabatic sound speed could differ greatly from one.
This could allow for an observable tensor to scalar ratio, where standard models of scalar driven inflation starting near the string scale seem to generically
predict an unobservable spectrum \cite{Kachru}.
This is work in progress.

\begin{figure}[t]
\includegraphics[width=8.2cm]{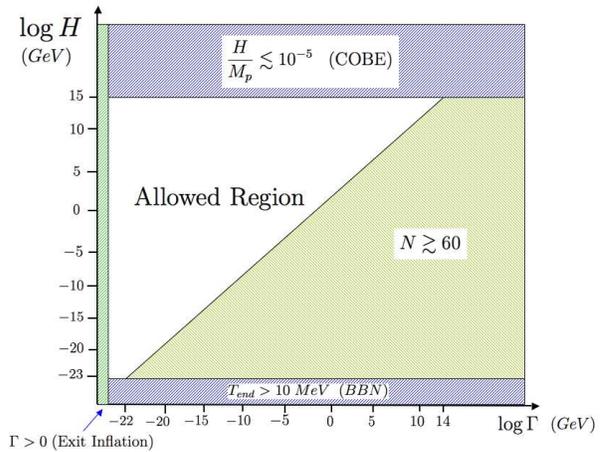}
\caption{\label{fig4} The figure above summarizes the constraints found on the Hubble scale during inflation and transition rate $\Gamma$ between 
levels.  In addition to these constraints one should also add the requirement that the transitions proceed via second order or weakly first order phase transitions.}
\end{figure}

\section{Further Considerations and Conclusions \label{section4}}
In this paper we have considered a cascading model for the early universe that provides a period of cosmological acceleration, which can account for the required number of efoldings. 
As the universe cascades, vacuum energy is converted into radiation inhomogeneously, resulting in a nearly scale invariant spectrum of cosmological density perturbations and
a small amount of gravitational waves. 
Once the radiation density overtakes the decaying vacuum energy, the model naturally exits in a radiation dominated universe 
with a temperature which we found can be as large as $T_r \approx 10^{15} GeV$.
As the universe evolves through the radiation and matter epochs the vacuum density will once again dominate the energy density if the decay does not proceed to zero vacuum energy. 

We have seen that our approach has one basic (in principle calculable) parameter,
the level decay rate $\Gamma$.  The number of e-foldings, the reheating,
and the density fluctuations all depend on $\Gamma$, and we find there does
exist a range of values of $\Gamma$ consistent with the data for all of
these, which might not have happened.

Although these preliminary findings are promising, much remains to be addressed.
A particularly pressing issue is a concrete derivation of the decay rate $\Gamma$ 
or equivalently a better understanding of the level spacing and the time spent in a given energy (density) level.
In fact, we argued in Section \ref{section4} that if $\Gamma$ is not taken constant, the result is a varying adiabatic sound speed which 
can result in density perturbations and gravity waves that would further distinguish the cascading model presented here from usual slow-roll inflation.

\acknowledgements
We would like to thank Gian Alberghi, Niayesh Afshordi, Robert Brandenberger, Ram Brustein, Juan Garcia-Bellido, Sera Cremonini, Fabio Finelli, Katie Freese, Bob Holdom, Nemanja Kaloper, Matt Kleban, Eugene Lim, Laura Mersini, Sergey Prokushkin, Jesse Thaler, and Henry Tye for useful discussions.
SW would like to thank Bologna University, the Galileo Galilei Institute for Theoretical Physics, and INFN for hospitality and partial support during the completion of this work.  SW was also supported in part by the National Science and Engineering Research Council of Canada.

%%%%%%%%%%%%%%%%     Bibliography    %%%%%%%%%%%%%%%%%%%%%%%

\end{document}